\documentclass[aps,twocolumn,floatfix,superscriptaddress,longbibliography]{revtex4}%
\usepackage[utf8]{inputenc}
\usepackage{bm}
\usepackage{times}
 \usepackage{algorithm,algorithmic}

\usepackage{mathrsfs} 
\usepackage{mathtools}
\usepackage{amsmath}
\usepackage[shortlabels]{enumitem}

\usepackage{graphicx,epic,eepic,epsfig,amsmath,latexsym,amssymb,verbatim,color}

\usepackage{amsfonts}       
\usepackage{nicefrac}       

\usepackage{amsmath}
\usepackage{bbm} 

\usepackage{float}
\usepackage{tikz}
\usetikzlibrary{chains}
\usetikzlibrary{fit}
\usepackage{pgflibraryarrows}		
\usepackage{pgflibrarysnakes}		

\usepackage{epsfig}
\usetikzlibrary{shapes.symbols,patterns} 
\usepackage{pgfplots}

\usepackage[strict]{changepage}
\usepackage{hyperref}
\hypersetup{colorlinks=true,citecolor=blue,linkcolor=blue,filecolor=blue,urlcolor=blue,breaklinks=true}

\usepackage[marginal]{footmisc}
\usepackage{url}
\usepackage{theorem}

\def\squareforqed{\hbox{\rlap{$\sqcap$}$\sqcup$}}
\def\qed{\ifmmode\squareforqed\else{\unskip\nobreak\hfil
\penalty50\hskip1em\null\nobreak\hfil\squareforqed
\parfillskip=0pt\finalhyphendemerits=0\endgraf}\fi}
\def\endenv{\ifmmode\;\else{\unskip\nobreak\hfil
\penalty50\hskip1em\null\nobreak\hfil\;
\parfillskip=0pt\finalhyphendemerits=0\endgraf}\fi}

\newcounter{remark}

\newcounter{example}

\mathchardef\ordinarycolon\mathcode`\:
\mathcode`\:=\string"8000
\def\vcentcolon{\mathrel{\mathop\ordinarycolon}}
\begingroup \catcode`\:=\active
  \lowercase{\endgroup
  \let :\vcentcolon
  }

\usepackage{cleveref}
\usepackage{graphicx}
\usepackage{xcolor}

\RequirePackage[framemethod=default]{mdframed}
\newmdenv[skipabove=7pt,
skipbelow=7pt,
backgroundcolor=darkblue!15,
innerleftmargin=5pt,
innerrightmargin=5pt,
innertopmargin=5pt,
leftmargin=0cm,
rightmargin=0cm,
innerbottommargin=5pt,
linewidth=1pt]{tBox}

\newmdenv[skipabove=7pt,
skipbelow=7pt,
backgroundcolor=blue2!25,
innerleftmargin=5pt,
innerrightmargin=5pt,
innertopmargin=5pt,
leftmargin=0cm,
rightmargin=0cm,
innerbottommargin=5pt,
linewidth=1pt]{dBox}
\newmdenv[skipabove=7pt,
skipbelow=7pt,
backgroundcolor=darkkblue!15,
innerleftmargin=5pt,
innerrightmargin=5pt,
innertopmargin=5pt,
leftmargin=0cm,
rightmargin=0cm,
innerbottommargin=5pt,
linewidth=1pt]{sBox}
\definecolor{darkblue}{RGB}{0,76,156}
\definecolor{darkkblue}{RGB}{0,0,153}
\definecolor{blue2}{RGB}{102,178,255}
\definecolor{darkred}{RGB}{195,0,0}

\newcommand{\nc}{\newcommand}
\nc{\rnc}{\renewcommand}
\nc{\beg}{\begin{equation}}
\nc{\eeq}{{\end{equation}}}
\nc{\beqa}{\begin{eqnarray}}
\nc{\eeqa}{\end{eqnarray}}
\nc{\lbar}[1]{\overline{#1}}
\nc{\bra}[1]{\langle#1|}
\nc{\ket}[1]{|#1\rangle}
\nc{\ketbra}[2]{|#1\rangle\!\langle#2|}
\nc{\braket}[2]{\langle#1|#2\rangle}

\nc{\proj}[1]{| #1\rangle\!\langle #1 |}
\nc{\avg}[1]{\langle#1\rangle}
\nc{\rank}{\operatorname{Rank}}
\nc{\smfrac}[2]{\mbox{$\frac{#1}{#2}$}}
\nc{\tr}{\operatorname{Tr}}
\nc{\ox}{\otimes}
\nc{\1}{\mathbbm{1}}
\nc{\dg}{\dagger}
\nc{\dn}{\downarrow}
\nc{\cA}{{\cal A}}
\nc{\cB}{{\cal B}}
\nc{\cC}{{\cal C}}
\nc{\cD}{{\cal D}}
\nc{\cE}{{\cal E}}
\nc{\cF}{{\cal F}}
\nc{\cG}{{\cal G}}
\nc{\cH}{{\cal H}}
\nc{\cI}{{\cal I}}
\nc{\cJ}{{\cal J}}
\nc{\cK}{{\cal K}}
\nc{\cL}{{\cal L}}
\nc{\cM}{{\cal M}}
\nc{\cN}{{\cal N}}
\nc{\cO}{{\cal O}}
\nc{\cP}{{\cal P}}
\nc{\cQ}{{\cal Q}}
\nc{\cR}{{\cal R}}
\nc{\cS}{{\cal S}}
\nc{\cT}{{\cal T}}
\nc{\cU}{{\cal U}}
\nc{\cV}{{\cal V}}
\nc{\cX}{{\cal X}}
\nc{\cY}{{\cal Y}}
\nc{\cZ}{{\cal Z}}
\nc{\cW}{{\cal W}}
\nc{\csupp}{{\operatorname{csupp}}}
\nc{\qsupp}{{\operatorname{qsupp}}}
\nc{\var}{{\operatorname{var}}}
\nc{\rar}{\rightarrow}
\nc{\lrar}{\longrightarrow}
\nc{\polylog}{{\operatorname{polylog}}}
\nc{\wt}{{\operatorname{wt}}}
\nc{\av}[1]{{\left\langle {#1} \right\rangle}}
\nc{\supp}{{\operatorname{supp}}}

\nc{\argmin}{{\operatorname{argmin}}}

\def\a{\alpha}
\def\b{\beta}

\def\t{\theta}

\def\x{\xi}

\nc{\RR}{{{\mathbb R}}}
\nc{\CC}{{{\mathbb C}}}
\nc{\FF}{{{\mathbb F}}}
\nc{\NN}{{{\mathbb N}}}
\nc{\ZZ}{{{\mathbb Z}}}
\nc{\PP}{{{\mathbb P}}}
\nc{\QQ}{{{\mathbb Q}}}
\nc{\UU}{{{\mathbb U}}}
\nc{\EE}{{{\mathbb E}}}
\nc{\id}{{\operatorname{id}}}

\nc{\CHSH}{{\operatorname{CHSH}}}

\nc{\be}{\begin{equation}}
\nc{\ee}{{\end{equation}}}
\nc{\bea}{\begin{eqnarray}}
\nc{\eea}{\end{eqnarray}}
\nc{\<}{\langle}
\rnc{\>}{\rangle}
\nc{\rU}{\mbox{U}}

\nc{\ob}[1]{#1}

\nc{\SEP}{{\text{\rm SEP}}}
\nc{\NS}{{\text{\rm NS}}}
\nc{\LOCC}{{\text{\rm LOCC}}}
\nc{\PPT}{{\text{\rm PPT}}}
\nc{\EXT}{{\text{\rm EXT}}}
\nc{\Sym}{{\operatorname{Sym}}}

\nc{\ERLO}{{E_{\text{r,LO}}}}
\nc{\ERLOCC}{{E_{\text{r,LOCC}}}}
\nc{\ERPPT}{{E_{\text{r,PPT}}}}
\nc{\ERLOCCinfty}{{E^{\infty}_{\text{r,LOCC}}}}
\nc{\Aram}{{\operatorname{\sf A}}}

\usepackage{tikz}
\usepackage{hyperref}
\hypersetup{colorlinks=true,citecolor=blue,linkcolor=blue,filecolor=blue,urlcolor=blue,breaklinks=true}

\makeatletter
\def\grd@save@target#1{%
  \def\grd@target{#1}}
\def\grd@save@start#1{%
  \def\grd@start{#1}}
\tikzset{
  grid with coordinates/.style={
    to path={%
      \pgfextra{%
        \edef\grd@@target{(\tikztotarget)}%
        \tikz@scan@one@point\grd@save@target\grd@@target\relax
        \edef\grd@@start{(\tikztostart)}%
        \tikz@scan@one@point\grd@save@start\grd@@start\relax
        \draw[minor help lines,magenta] (\tikztostart) grid (\tikztotarget);
        \draw[major help lines] (\tikztostart) grid (\tikztotarget);
        \grd@start
        \pgfmathsetmacro{\grd@xa}{\the\pgf@x/1cm}
        \pgfmathsetmacro{\grd@ya}{\the\pgf@y/1cm}
        \grd@target
        \pgfmathsetmacro{\grd@xb}{\the\pgf@x/1cm}
        \pgfmathsetmacro{\grd@yb}{\the\pgf@y/1cm}
        \pgfmathsetmacro{\grd@xc}{\grd@xa + \pgfkeysvalueof{/tikz/grid with coordinates/major step}}
        \pgfmathsetmacro{\grd@yc}{\grd@ya + \pgfkeysvalueof{/tikz/grid with coordinates/major step}}
        \foreach \x in {\grd@xa,\grd@xc,...,\grd@xb}
        \node[anchor=north] at (\x,\grd@ya) {\pgfmathprintnumber{\x}};
        \foreach \y in {\grd@ya,\grd@yc,...,\grd@yb}
        \node[anchor=east] at (\grd@xa,\y) {\pgfmathprintnumber{\y}};
      }
    }
  },
  minor help lines/.style={
    help lines,
    step=\pgfkeysvalueof{/tikz/grid with coordinates/minor step}
  },
  major help lines/.style={
    help lines,
    line width=\pgfkeysvalueof{/tikz/grid with coordinates/major line width},
    step=\pgfkeysvalueof{/tikz/grid with coordinates/major step}
  },
  grid with coordinates/.cd,
  minor step/.initial=.2,
  major step/.initial=1,
  major line width/.initial=2pt,
}
\makeatother

\usepackage{thmtools}
\usepackage{thm-restate}
\usepackage{etoolbox}
\makeatletter
\def\problem@s{}
\newcounter{problems@cnt}

\newcommand{\allproblems}{\problem@s}
\makeatother

\definecolor{beamer}{rgb}{0.2,0.2,0.7}
\definecolor{colorone}{rgb}{1,0.36,0.03}
\definecolor{colortwo}{rgb}{0.4,0.77,0.17}
\definecolor{colorthree}{rgb}{0.01,0.51,0.93}
\definecolor{colorfour}{rgb}{0.47,0.26,0.58}
\definecolor{colorfive}{rgb}{0.12,0.55,0.16}
\usepackage{tcolorbox}
\usepackage{relsize}
\usepackage{graphicx}
\nc{\st}{\text{subject to} \ }
\nc{\supre}{\text{supremum} \ }
\nc{\sdp}{\text{sdp}}
\usepackage[qm]{qcircuit}
\usepackage{array}

\newcommand{\KW}[1]{\textcolor{beamer}{(KW: #1)}}
\newcommand{\zx}[1]{\textcolor{darkred}{(ZX: #1)}}
\newcommand{\xq}[1]{\textcolor{cyan}{(XQ: #1)}}

\begin{document}
\title{Detecting and quantifying entanglement on near-term quantum devices}
\author{Kun Wang}
\thanks{wangkun28@baidu.com}
\author{Zhixin Song}
\author{Xuanqiang Zhao}
\author{Zihe Wang}
\author{Xin Wang}
\thanks{wangxin73@baidu.com}
\affiliation{Institute for Quantum Computing, Baidu Research, Beijing 100193, China}

\begin{abstract}
Quantum entanglement is a key resource in quantum technology, and its quantification is a vital task in the current Noisy Intermediate-Scale Quantum (NISQ) era. This paper combines hybrid quantum-classical computation and quasi-probability decomposition to propose two variational quantum algorithms, called Variational Entanglement Detection (VED) and Variational Logarithmic Negativity Estimation (VLNE), for detecting and quantifying entanglement on near-term quantum devices, respectively. VED makes use of the positive map criterion and works as follows. Firstly, it decomposes a positive map into a combination of quantum operations implementable on near-term quantum devices. It then variationally estimates the minimal eigenvalue of the final state, obtained by executing these implementable operations on the target state and averaging the output states. Deterministic and probabilistic methods are proposed to compute the average. At last, it asserts that the target state is entangled if the optimized minimal eigenvalue is negative. VLNE builds upon a linear decomposition of the transpose map into Pauli terms and the recently proposed trace distance estimation algorithm. It variationally estimates the well-known logarithmic negativity entanglement measure and could be applied to quantify entanglement on near-term quantum devices. Experimental and numerical results on the Bell state, isotropic states, and Breuer states show the validity of the proposed entanglement detection and quantification methods.
\end{abstract}

\date{\today}
\maketitle

\section{Introduction}
It is widely believed that we are now in the Noisy Intermediate Scale Quantum (NISQ) era~\cite{preskill2018quantum}, where quantum computers with $50$-$100$ qubits are available
while noise in quantum gates severely limits the quantum circuits that can be executed reliably.
It thus becomes important to make the best use of today's NISQ devices to design practical applications. 
One promising scheme for near-term quantum applications is the variational quantum algorithms (VQA)~\cite{mcclean2016theory}, which have been applied to solve many tasks including Hamiltonian ground and excited states preparation~\cite{peruzzo2014variational,nakanishi2019subspace}, quantum state distance estimation~\cite{cerezo2020variational,Chen2020a}, quantum metrology~\cite{Beckey2020,Koczor2019a}, and quantum data compression~\cite{Romero2017,Wan2017,Cao2020}. These variational quantum algorithms involve evaluating and optimizing loss functions that depend on parameters in parameterized quantum circuits (PQC). They are regarded as well-suited for execution on NISQ devices by combining quantum computers with classical computers. We refer the readers
to~\cite{Endo2020,Cerezo2020} for a detailed review on VQA.

Quantum entanglement~\cite{horodecki2009quantum},
the most nonclassical manifestation of quantum mechanics, 
has been identified as invaluable resource enabling a tremendous number of tasks ranging from quantum information processing~\cite{bennett1992communication,bennett1993teleporting},
quantum cryptography~\cite{bell1964einstein,clauser1970proposed,ekert1991quantum},
quantum algorithms~\cite{shor1994algorithms,jozsa1997entanglement,childs2010quantum},
quantum communication~\cite{bennett1999entanglement,bennett2002entanglement},
to measurement-based quantum computing~\cite{raussendorf2001one,raussendorf2003measurement,gross2007novel,briegel2009measurement}.
As so, the ability to manipulate quantum entanglement is the cornerstone to 
achieve real applications of quantum technologies. 
A number of theoretical and experimental methods have been proposed 
in the past $20$ years for entanglement detection and quantification~\cite{horodecki2009quantum,guhne2009entanglement,Friis2019}.
For example, entanglement can be detected 
via entanglement witnesses~\cite{horodecki1996Separability,guhne2006nonlinear}, 
Bell's inequalities~\cite{bell2004speakable}, 
quantum Fisher information~\cite{pezze2009entanglement},
realignment criterion~\cite{chen2002matrix,rudolph2005further}, range criterion~\cite{horodecki1997separability},
and majorization criterion~\cite{nielsen2001separable}, to name a few. These methods commonly assume that prior information about the target state is known. A direct way to obtain such information is to perform quantum state tomography and reconstruct the density matrix~\cite{d2003quantum,steffen2006measurement}.
However, tomography becomes unrealistic as the number of required measurement settings scales exponentially with the size of the system. Briefly speaking, though there are many methods proposed for detecting and quantifying quantum entanglement, they are not specially designed for near-term quantum devices and thus are not directly applicable in most cases, rendering reliable detection and quantification of quantum entanglement on near-term quantum devices a vital challenge.

In this paper, we combine VQA and the quasi-probability decomposition technique~\cite{Buscemi2013,Buscemi2014,pashayan2015estimating,temme2017error,endo2018practical,takagi2020optimal,jiang2020physical} 
to propose the Variational Entanglement Detection (VED) and Variational Logarithmic Negativity Estimation (VLNE) algorithms, aiming to detect and quantify quantum entanglement on near-term quantum devices, respectively. VED uses criteria based on positive maps as a bridge and works as follows. Given an unknown target bipartite quantum state,  it firstly decomposes the chosen positive map into a linear combination of NISQ implementable quantum operations. Then, it variationally estimates the minimal eigenvalue of the final state, which is obtained by executing these quantum operations on the target state and averaging the output states. Two methods are proposed to compute the average: the first one averages the output states according to the quasi-probability distribution, and the second one estimates the average via the sampling technique and is probabilistic. At last, it asserts that the target state is entangled if the optimized minimal eigenvalue is negative. Following the idea of VED, VLNE variationally computes the well-known log-negativity entanglement measure, building on a linear decomposition of the transpose map into Pauli terms and the recently proposed trace distance estimation algorithm.
Our main contributions can be summarized as follows:
\begin{enumerate}
    \item We combine VQA and the quasi-probability decomposition technique to propose the VED framework, contributing a feasible solution for detecting entanglement on near-term quantum devices.
    \item We combine VQA and the quasi-probability decomposition technique to propose the VLNE algorithm that could estimate the well-known logarithmic negativity~\cite{plenio2005logarithmic}, which may lead to various applications in near-term quantum information processing.
\end{enumerate}
Experimental and numerical results reveal the validity of the proposed entanglement detection and quantification methods.

Our paper is structured as follows.
In Sec.~\ref{sec:preliminaries}, we set the notations and 
briefly summarize the entanglement criteria based on positive maps. In Sec.~\ref{sec:VED}, we present our first main result: the VED framework. In Sec.~\ref{sec:pm}, we elaborate on three prominent positive maps to illustrate how the VED framework is applied. In Sec.~\ref{sec:VEQ}, we present our second main result: the VLNE framework, which estimates the logarithmic negativity entanglement measure. In Secs.~\ref{sec:ibmq-santiago} and~\ref{sec:numercs}, experiments on \href{https://quantum-computing.ibm.com}{IBM-Q} and numerical simulations on  \href{https://github.com/PaddlePaddle/Quantum}{Paddle Quantum} are conducted on various bipartite quantum states of interests to show the validity of the proposed methods. We finally conclude in Sec.~\ref{sec:conclusion}.

\section{Preliminaries}\label{sec:preliminaries}

\subsection{Notations}

In this section, we set the notation and define several quantities that will be used throughout this paper. We will frequently use symbols such as $\mathcal{H}_A$ and $\mathcal{H}_B$ to denote Hilbert spaces associated with quantum systems $A$ and $B$, respectively. We use $d_A$ to denote the dimension of the system $A$. The set of linear operators acting on $A$ is denoted by $\cL(\mathcal{H}_A)$. We write an operator with subscript to indicate the system that the operator acts on, e.g., $X_{AB}$, and write $X_A:=\tr_B X_{AB}$. For a linear operator $X\in\cL(\mathcal{H}_A)$, we define its modulus $|X|:=\sqrt{X^\dagger X}$, where $X^\dagger$ is the adjoint operator of $X$. The trace norm of $X$ is defined as $\|X\|_1 :=\tr |X|$. We use $X\geq0$ to indicate that $A$ is positive semidefinite. A quantum map $\cN$ that transforms linear operators to linear operators in the system $A$ is abbreviated as $\cN_{A\to A}$. We use the calligraphic letters (e.g., $\cN$, $\cR$, and $\cO$) to represent linear quantum maps and use $\id_A$ to represent the identity map on system $A$. We say $\cN_{A\to A}$ is trace-preserving if $\tr[\cN(X)]=\tr[X]$ for arbitrary $X\in\cL(\cH_A)$,
is {positive} if $\cN(X)\geq0$ for arbitrary $X\geq0$,
and is {completely positive} if $\id_R\ox\cN$ is positive for arbitrary reference system $R$.

Given a Hermitian operator $X$ in system $A$, we
denote by $\lambda_{\min}(X)$ its minimal eigenvalue.
We have the following variational characterization:
\begin{align}\label{eq:min-eigenvalue}
    \lambda_{\min}(X) = \min_{\ket{\psi}}\bra{\psi} X\ket{\psi},
\end{align}
where the minimization ranges over the set of pure states on system $A$.

We use $\mathbb{R}$ to represent the real field.
We introduce the sign function $\operatorname{sgn}:\mathbb{R}\to\{\pm1,0\}$ 
as $\forall x<0$, $\operatorname{sgn}(x)=-1$;
$\forall x>0$, $\operatorname{sgn}(x)=1$; 
and $\operatorname{sgn}(0)=0$. 
All logarithms are in base $2$ in this paper.

\subsection{Pauli operators and Pauli channels}

The four Pauli operators in the qubit space are defined as
\begin{subequations}\label{eq:Pauli}
\begin{alignat}{3}
  I \equiv \sigma_0 &:= \begin{pmatrix} 1 & 0 \\ 0 & 1\end{pmatrix},
   && \quad &
  X \equiv \sigma_1 &:= \begin{pmatrix} 0 & 1 \\ 1 & 0\end{pmatrix}, \\
  Y \equiv \sigma_2 &:= \begin{pmatrix} 0 & -i \\ i & 0\end{pmatrix},
   && \quad &
  Z \equiv \sigma_3 &:= \begin{pmatrix} 1 & 0 \\ 0 & -1\end{pmatrix}.
\end{alignat}
\end{subequations}
They provide a basis for the qubit linear operators, i.e.,
arbitrary qubit linear operator can be decomposed w.r.t. 
this basis.
For the $n$-qubit case, one can construct a set of Pauli operators, 
which we call the Pauli set, as
\begin{align}\label{eq:Pauli-set}
\bm{P}_n := \left\{\bigotimes_{k=1}^n\sigma_{q_k}
        \;\middle\vert\; q_k = 0,1,2,3\right\}
            \equiv \left\{I, X, Y, Z\right\}^{\ox n}.
\end{align}
The $n$-qubit Pauli set has size $\vert\bm{P}_n\vert=4^n$. 
Note that $\bm{P}_n$ forms a basis for the $n$-qubit linear operators.
One can see from~\eqref{eq:Pauli-set} that
each Pauli operator $P \in \bm{P}_n$ can be represented uniquely by a \emph{quaternary} sequence:
\begin{align}\label{eq:Pauli-quaternary}
    \bm{q} = q_1\cdots q_n, \quad \text{with} \quad q_k = 0, 1,2,3.
\end{align}
As so, we use $P_{\bm{q}}$ to represent the $n$-qubit Pauli operator
that is uniquely determined by the sequence $\bm{q}$.
Since each qubit Pauli operator is unitary, so is each 
$n$-qubit Pauli operator, due to the construction.
Given a Pauli operator $P \in \bm{P}_n$, we denote by
$\cP(\cdot) := P(\cdot)P^\dagger$ its induced Pauli channel.

\subsection{Entanglement detection via positive maps}

Let $\rho_{AB}$ be a bipartite quantum state in the composite system $AB$. 
By definition $\rho_{AB}$ is \emph{separable} if it can be decomposed 
into a convex combination of tensor products of states describing local systems as~\cite{werner1989quantum}
\begin{align}\label{eq:separable}
\rho_{AB} = \sum_x p_x \proj{\psi_x}_A\ox\proj{\phi_x}_B,
\end{align}
where $p_x\geq 0$, $\sum_xp_x=1$, and $\{\ket{\psi_x}\}_x$ and $\{\ket{\phi_x}\}_x$
are two sets of pure states in systems $A$ and $B$, respectively.
Otherwise, $\rho_{AB}$ is \emph{entangled}.
Given the definition, it is natural to ask whether 
a given unknown bipartite quantum state is separable or entangled, known
as the \emph{separability problem}.
This problem has been shown to be NP-hard~\cite{gurvits2003classical,gharibian2008strong}.
There are many separability criteria that have been proposed to
determine the separability or entanglement of bipartite quantum states 
as necessary conditions~\cite{guhne2009entanglement,horodecki2009quantum}.

One of the most celebrated criteria for distinguishing separable states from
entangled states are the positive map criterion.
The core of the positive map criterion is that one subjects a subsystem of $\rho_{AB}$ 
to a positive (but not completely positive) map $\cN_{B\to B}$
that preserves the positivity of inputs. 
If $\rho_{AB}$ is a product state, i.e.,
it is of the form $\rho_A\ox\rho_B$, the resulting operator $\rho_A\ox\cN(\rho_B)$ is still positive.
Consequently, due to the linearity, an arbitrary separable state is mapped into
some positive operator by this map. On the other hand, if $\rho_{AB}$ is entangled,
the output operator $\cN_{B\to B}(\rho_{AB})$ may be no longer positive; the transpose map is a
prominent example~\cite{peres1996separability}.
That is to say, the negative spectrum of the output operator indicates entanglement of the input state.
Mathematically, the positive map criterion states that 
a bipartite quantum state $\rho_{AB}$ is separable if and only if
for arbitrary system $C$ and arbitrary positive (but not completely positive)
map $\cN_{B\to C}$, it holds that $\cN_{B\to C}(\rho_{AB})\geq 0$~\cite{horodecki1996Separability}.

Despite its proven efficiency in entanglement detection, the positive map criterion is not directly applicable in practice, especially on recent NISQ devices. This is an immediate consequence of the fact that generically positive but not completely positive maps do not represent physically implementable quantum operations~\cite{kraus1983states}  and thus cannot be realized in near-term quantum devices. In the following, we show how to overcome this obstacle and employ the positive map criterion to detect entanglement on NISQ devices.

\section{Quantum entanglement detection}\label{sec:VED}

In this section, we integrate variational quantum algorithms with
the quasi-probability decomposition technique~\cite{Buscemi2013,Buscemi2014,pashayan2015estimating,temme2017error,endo2018practical,takagi2020optimal,jiang2020physical} to propose 
a bipartite entanglement detection framework specially designed for 
near-term quantum devices, 
using positive map criterion as a bridge. 
For simplicity, we assume $A$ and $B$ are two $n$-qubit quantum systems 
throughout this section. However, we remark that the proposed framework can applied
to bipartite systems with different dimensions directly.

Let $\Delta$ be a discrete set of quantum operations that are implementable in the near-term 
quantum devices. For example, one may choose $\Delta$ to be the set of implementable operations
introduced in~\cite{endo2018practical,takagi2020optimal}. 
Alternatively, one may set $\Delta$ to be the set of 
Pauli channels induced by Pauli operators from
the Pauli set~\eqref{eq:Pauli-set}.
For a positive (but not completely positive) and trace-preserving map $\cN_{B\to B}$, 
we assume that it can be decomposed w.r.t. $\Delta$ as
\begin{align}\label{eq:decomposition}
    \cN(\cdot) = \sum_{\cO\in\Delta} r_{\cO} \cO(\cdot),\; r_{\cO}\in\mathbb{R}.
\end{align}
Note that such a decomposition always exists if $\Delta$ 
contains a universal basis~\cite{endo2018practical}.
The trace-preserving condition imposes $\sum_{\cO}r_{\cO}=1$.
We emphasize that there must exist negative coefficients $r_{\cO}$ 
since otherwise, $\cN$ is completely positive.
Given a bipartite quantum state $\rho_{AB}$, we have
\begin{align}\label{eq:decomposition-2}
  \sigma_{AB} := \cN_{B\to B}(\rho_{AB})
= \sum_{\cO\in\Delta} r_{\cO} \cO_{B\to B}(\rho_{AB}).
\end{align}
To see if $\rho_{AB}$ can be detected by $\cN$, i.e.,
if $\rho_{AB}$ is entangled from $\cN$'s perspective, 
we need to check if the output state $\sigma_{AB}$ has a negative eigenvalue or not.
Denote by $\lambda_{\min}(\sigma_{AB})$ as the smallest eigenvalue of $\sigma_{AB}$. 
By the positive map criterion, if $\rho_{AB}$ is separable, 
then it must hold that $\lambda_{\min}(\sigma_{AB})\geq0$. 
Equivalently, if $\lambda_{\min}(\sigma_{AB})<0$, 
we safely conclude that $\rho_{AB}$ is entangled and 
it can be detected by the positive map $\cN_{B\to B}$. 
This highlights the importance of computing or 
estimating $\lambda_{\min}(\sigma_{AB})$ in entanglement detection.

\subsection{Deterministic detection}

\begin{figure}
    \centering
    \includegraphics[width=0.4\textwidth]{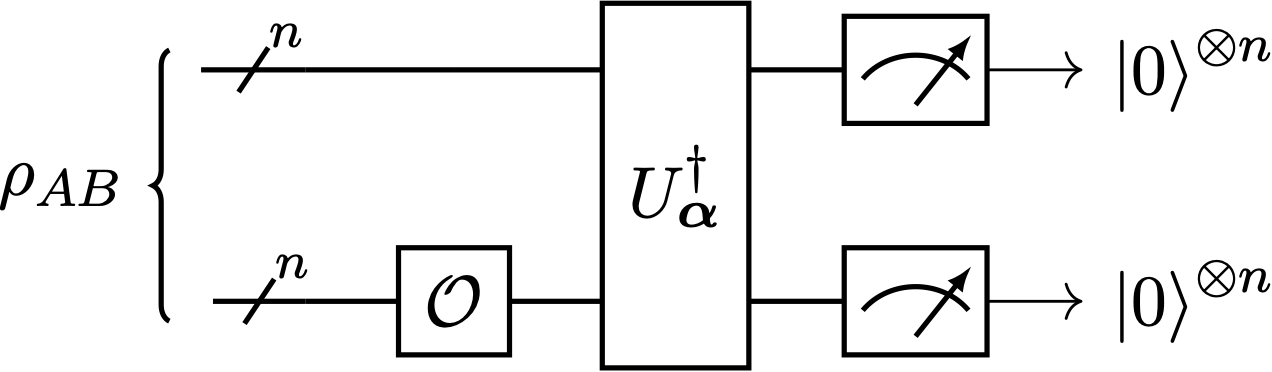}
    \caption{The simplified quantum circuit that estimates 
            the overlap $\bra{\psi} \cO(\rho_{AB}) \ket{\psi}$ in~\eqref{eq:tmp3} 
            for a given implementable operation $\cO$,
            where $\ket{\psi}:=U_{\bm\a}\ket{0}^{\ox 2n}$ is the parameterized
            input state.}
    \label{fig:overlap}
\end{figure}

As we have argued, $\sigma_{AB}$ cannot be obtained directly via $\cN(\rho)$
since $\cN$ does not represent physically implementable quantum operations.
Fortunately, the decomposition~\eqref{eq:decomposition-2}
empowers us an effective way to simulate the role of $\cN$ 
and reconstruct $\sigma_{AB}$ as an average of a set of output states, 
obtained using quantum circuits implementable in near-term devices. 
This decomposition technique, combined with the variational quantum algorithm, 
enables a general framework that estimates $\lambda_{\min}(\sigma_{AB})$, 
whose value can witness the entanglement of the input state $\rho_{AB}$. 
We call this framework the \emph{Variational Entanglement Detection (VED)}. 
The core idea is to use the linear decomposition~\eqref{eq:decomposition-2} of 
the target state $\sigma_{AB}$ and the framework goes as follows.
First of all, by~\eqref{eq:min-eigenvalue} it holds that
\begin{align}
    \lambda_{\min}(\sigma_{AB}) 
&=  \min_{\ket{\psi}_{AB}} \bra{\psi} \sigma_{AB} \ket{\psi}\label{eq:tmp1}  \\
&=  \min_{\ket{\psi}_{AB}} 
    \sum_{\cO\in\Delta} r_\cO \bra{\psi} \cO(\rho_{AB}) \ket{\psi},\label{eq:tmp2}
\end{align}
where the minimization ranges over all pure bipartite quantum states $\ket{\psi}_{AB}$ in $AB$.
We use a variational quantum circuit with parameters $\bm{\a}$ 
to prepare the test state $\ket{\psi}$.
More precisely, we choose a parametrized quantum circuit ansatz that generates
a unitary $U(\bm{\a})$ and prepare the test state via $\ket{\psi} = U(\bm{\a})\ket{0}^{\ox 2n}$.
Each inner product $\bra{\psi} \cO(\rho) \ket{\psi}$ 
in~\eqref{eq:tmp2} can be estimated via the 
canonical Swap Test subroutine~\cite{buhrman2001quantum}, 
as both $U(\bm{\a})$ and $\cO$ can be implemented in near-term devices.
However, this subroutine costs a total number of $4n+1$ qubits and requires a 
$4n$-qubit SWAP gate, which is resource consuming when $n$ becomes large.
Here we explore the special structure of the overlap $\bra{\psi} \cO(\rho) \ket{\psi}$ 
and propose an qubit efficient estimating procedure which uses $2n$ qubits 
and avoids the use of expensive SWAP gate. First of all, notice that
\begin{align}
&\;\bra{\psi(\bm\a)} \cO(\rho_{AB})\ket{\psi(\bm\a)} \\
=&\; \tr\left[
    \cO(\rho_{AB})U_{\bm\a}\ket{0^{2n}}\bra{0^{2n}}U_{\bm\a}^\dagger\right] \\
=&\; \bra{0^{2n}}U_{\bm\a}^\dagger \cO(\rho_{AB}) U_{\bm\a}\ket{0^{2n}},\label{eq:tmp3}
\end{align}
where the second equality follows from the cyclic property of trace function.
Since each $\cO$ is implementable on near-term quantum devices,
we may use $\rho_{AB}$ as input to the quantum circuit implementing $\cO$,
and estimate the overlap $\bra{\psi} \cO(\rho_{AB}) \ket{\psi}$
using the quantum circuit illustrated in Fig.~\ref{fig:overlap}.
The overlap is obtained by counting the relative frequency of the measurement
outcome $0^{2n}$.
Then, we repeat the estimation procedure $\vert\Delta\vert$ times, where
$\vert\Delta\vert$ is the size of $\Delta$,
to obtain the overlaps for different $\cO$ in~\eqref{eq:tmp2}.
With these data in hand, we compute the following loss function
\begin{align}\label{eq:loss}
    L(\bm\a) := \sum_{\cO\in\Delta} 
            r_\cO \bra{\psi(\bm\a)} \cO(\rho_{AB}) \ket{\psi(\bm\a)}.
\end{align}
At last, we perform gradient-based optimization methods including SGD~\cite{kiefer1952stochastic} and Adam~\cite{kingma2014adam}
to minimize the loss function $L(\bm \alpha)$
by varying the parameters $\bm{\a}$,
whose value will determine the separability of the input state $\rho_{AB}$.
More precisely, if $L(\bm \alpha)$ is negative, we conclude that $\rho_{AB}$ is entangled,
since by the positive map criterion, separable states cannot yield a negative spectrum.

Taking into account the noise in NISQ quantum devices, 
we may introduce a tolerance threshold $\delta>0$ 
so that $L(\bm\a)<-\delta$ implies the input state is entangled. 
This threshold $\delta$ can be set with prior knowledge about 
the noise characterization on the NISQ devices.
What's more, for the purpose of entanglement detection, 
it is unnecessary to minimize $L(\bm\a)$ since the condition $L(\bm\a)<0$ is 
sufficient to assert that the input state is entangled. 
Based on this observation, we can terminate the 
optimization procedure that minimizes the loss function $L(\alpha)$
in advance to save the optimization cost.
The detailed VED framework is summarized in Algorithm~\ref{alg:VED}.
We name it the deterministic VED to 
distinguish it from the probabilistic framework 
described in the next section.

\begin{algorithm}[H] 
\caption{Deterministic VED}
\begin{algorithmic}[1] \label{alg:VED}
\STATE Input: $2n$-qubit quantum state $\rho_{AB}$, 
        decomposition~\eqref{eq:decomposition} of the positive map $\cN$,
        parameterized quantum circuits $U(\bm\a)$ with initial parameters $\bm\a$, 
        and tolerance $\delta$;

\STATE Initialize $L(\bm\a) = 0$;

\FORALL{$\cO\in\Delta$ such that $r_{\cO}\neq0$} 
    \STATE\hskip0.5em  Apply $U_{\bm\a}$ to $\ket{0}^{\ox 2n}$ 
            and obtain test state $\ket{\psi} = U_{\bm\a}\ket{0}^{\ox 2n}$; 
            
    \STATE\hskip0.5em  Input $\rho_{AB}$ and
            compute the overlap $c_\cO := \bra{\psi} \cO(\rho_{AB}) \ket{\psi}$ 
            
            \hskip0.5em using the quantum circuit in Fig.~\ref{fig:overlap};
            
    \STATE\hskip0.5em Update the loss function $L(\bm\a) = L(\bm\a) + r_\cO c_\cO$,
          
          \hskip0.5em where $r_\cO$ is given by the decomposition~\eqref{eq:decomposition};  
\ENDFOR

\STATE Perform optimization methods to minimize $L(\bm\a)$; 
        terminate the optimization if  
        the error tolerance is satisfied: $L(\bm\a) < -\delta$;
\STATE Output "Entangled" if the optimized $L(\bm\a) < -\delta$.
\end{algorithmic}
\end{algorithm}

\subsection{Probabilistic detection}

In Algorithm~\ref{alg:VED}, we have used a brute-force approach, 
where we iterate over the set of implementable operations $\Delta$, 
to estimate the loss function $L(\bm\a)$.
Actually, $L(\bm\a)$ can be estimated in a probabilistic 
way using the sampling technique,
by virtue of the quasi-probability decomposition~\eqref{eq:decomposition}.
This new method would be beneficial when 
the number of decomposed operations in~\eqref{eq:decomposition} 
with non-zero coefficients is large while the sampling cost is relatively low.
Now we describe the sampling method accurately.
First of all, notice that the decomposition~\eqref{eq:decomposition} 
induces a quasi-probability distribution $\{r_{\cO}\}_{\cO\in\Delta}$ 
over $\Delta$.
From this quasi-probability distribution, 
we can construct a probability distribution $\{p_{\cO}\}_{\cO\in\Delta}$
using the canonical technique, i.e.,
\begin{align}\label{eq:distribution}
    p_{\cO} := \frac{\vert r_\cO\vert}{\gamma},\quad
    \gamma := \sum_{\cO\in\Delta}\vert r_{\cO}\vert.
\end{align}
Substituting~\eqref{eq:distribution} into~\eqref{eq:loss} yields
\begin{align}
    L(\bm\a) 
&= \gamma\sum_{\cO\in\Delta} \operatorname{sgn}(r_\cO) p_{\cO}\bra{\psi(\bm\a)} \cO(\rho_{AB}) \ket{\psi(\bm\a)} \\
&= \mathbb{E}_{\cO}
    \left[\gamma\operatorname{sgn}(r_\cO)\bra{\psi(\bm\a)} \cO(\rho_{AB}) \ket{\psi(\bm\a)}\right],
    \label{eq:expectation2}
\end{align}
where $\mathbb{E}(X)$ denotes the expectation of the random variable $X$,
and the expectation in~\eqref{eq:expectation2} is evaluated 
w.r.t. the probability distribution $\{p_{\cO}\}_{\cO\in\Delta}$.
Based on~\eqref{eq:expectation2}, we propose Algorithm~\ref{alg:VED-via-sampling},
which can be viewed as a probabilistic version of Algorithm~\ref{alg:VED}.
In particular, Algorithm~\ref{alg:VED-via-sampling} replaces the 
brute-force approach (steps 3-7) in Algorithm~\ref{alg:VED} with 
the sampling approach, yielding a probabilistic algorithm.

\begin{algorithm}[H] 
\caption{Probabilistic VED}
\begin{algorithmic}[1] \label{alg:VED-via-sampling}
\STATE Input: $2n$-qubit quantum state $\rho_{AB}$, 
        decomposition~\eqref{eq:decomposition} of the positive map $\cN$,
        parameterized quantum circuits $U(\bm\a)$ with initial parameters $\bm\a$, 
        error tolerance $\delta$, and fail probability $\varepsilon$.

\STATE Initialize $L'(\bm\a) = 0$;

\STATE Compute $\gamma$ defined in~\eqref{eq:distribution}
        and set $M = 2\gamma^2\log(2/\varepsilon)/\delta^2$; 
 
\FORALL{$m=1,\cdots,M$} 
    \STATE\hskip0.5em  Apply $U_{\bm\a}$ to $\ket{0}^{\ox 2n}$ 
            and obtain test state $\ket{\psi} = U_{\bm\a}\ket{0}^{\ox 2n}$; 
    
    \STATE\hskip0.5em  Sample a quantum operation $\cO^{(m)}$ from $\Delta$ 
    
            \hskip0.5em according to the probability distribution 
                        $\{p_{\cO}\}_{\cO\in\Delta}$ in~\eqref{eq:distribution};
            
            \hskip0.5em Let $r^{(m)}$ be the coefficient of $\cO^{(m)}$ in~\eqref{eq:decomposition};
            
    \STATE\hskip0.5em  Input $\rho_{AB}$ and compute the overlap 
                        $c^{(m)} := \bra{\psi} \cO^{(m)}(\rho) \ket{\psi}$ 
            
            \hskip0.5em using the quantum circuit in Fig.~\ref{fig:overlap};

    \STATE\hskip0.5em Compute 
            $L^{(m)} = \gamma\operatorname{sgn}(r^{(m)}) c^{(m)}$;
\ENDFOR

\STATE Compute the loss function $L'(\bm\a) = \frac{1}{M}\sum_{m=1}^M L^{(m)}$;  

\STATE Perform optimization methods to minimize $L'(\bm\a)$; 
        terminate the optimization if  
        the error tolerance is satisfied: $L'(\bm\a) < -\delta$;
\STATE Output "Entangled" if the optimized $L'(\bm\a) < -\delta$.
\end{algorithmic}
\end{algorithm}

Let's analyze Algorithm~\ref{alg:VED-via-sampling} in depth. 
First, we remark that the obtained $L'(\bm\a)$ in step 11 
of Algorithm~\ref{alg:VED-via-sampling} is an unbiased estimator 
of true value $L(\bm\a)$ due to~\eqref{eq:expectation2}. 
Second, since $\vert L^{(m)}\vert\leq\gamma$, 
we can apply the Hoeffding inequality~\cite{hoeffding1994probability} to ensure that
$M = 2\gamma^2\log(2/\varepsilon)/\delta^2$
number of samples would estimate 
the true value $L(\bm\a)$ within error $\delta$ 
with success probability no less than $1-\varepsilon$, i.e.,
\begin{align}\label{eq:error-probability}
p\left(\vert L'(\bm\a)  - L(\bm\a)\vert \leq \delta \right) \geq 1 - \varepsilon.
\end{align}
This confirms the validity 
of the sampling procedure (steps 4-9) of Algorithm~\ref{alg:VED-via-sampling}.
We call $\gamma$ the \emph{sampling cost} since it determines $M$, 
the number of samples required to achieve the desired precision.
At last, we examine the success probability of the algorithm, given
the success probability condition~\eqref{eq:error-probability} 
of the sampling procedure.
Assume the optimization procedure repeats $K$ times. 
The overall success probability of Algorithm~\ref{alg:VED-via-sampling}
is no less than $1- K\varepsilon$, 
as a direct corollary of~\eqref{eq:error-probability} and the union bound. 
That is to say, if Algorithm~\ref{alg:VED-via-sampling} outputs "Entangled",
$\rho_{AB}$ is entangled with probability larger than $1- K\varepsilon$.

To summarize, we have proposed two variational entanglement detection methods.
Algorithm~\ref{alg:VED} is \emph{deterministic} in the sense that
whenever it outputs "Entangled", one can safely assert 
that $\rho_{AB}$ is entangled. On the other hand, 
Algorithm~\ref{alg:VED-via-sampling} is \emph{probabilistic} in the sense that
even if it outputs "Entangled", one can only 
declare that $\rho_{AB}$ is entangled with certain success probability.
Nevertheless, when the number of decomposed operations in~\eqref{eq:decomposition}
with non-zero coefficients is large while the simulation cost
$\gamma$ is relatively low, the latter method may be beneficial.
In this case, one can reduce the number of iterations via sampling 
and thus save computational resources. 
Algorithm~\ref{alg:VED-via-sampling}
scarifies precision for efficiency in entanglement detection.

\section{Prominent positive maps}\label{sec:pm}

In Sec.~\ref{sec:VED} we have outlined the general deterministic and probabilistic 
VED frameworks for detecting entanglement via positive map criterion.
In this section, we elaborate on three prominent positive maps---the transpose map~\cite{peres1996separability},
the reduction map~\cite{Cerf1999}, 
and the enhanced reduction map~\cite{breuer2006optimal,hall2006new}---to 
illustrate how the deterministic VED framework works.
We choose the set of NISQ implementable 
quantum operations $\Delta$ 
to be the set of Pauli channels induced by Pauli operators 
from the Pauli set~\eqref{eq:Pauli-set}, i.e.,
\begin{align}\label{eq:Delta-Pauli}
  \Delta := \left\{ \cP \;\middle\vert\; \cP(\cdot) = P(\cdot) P^\dagger, P \in \bm{P}_n\right\}.
\end{align}
For each of the three positive maps under consideration, we firstly decompose it w.r.t. $\Delta$
as~\eqref{eq:decomposition} and then adopt the variational
 framework summarized in Algorithm~\ref{alg:VED}
to fulfill entanglement detection.
However, we remind that not all positive maps can be 
decomposed w.r.t. the set of Pauli channels. 

Here are remarks for the three criteria under consideration.
First, the reduction criterion is strictly weaker than both the transpose criterion and the enhanced reduction criterion, in the sense that the states that can be detected by the first criterion can also be detected by the latter two criteria. Second, there is no inclusion relation between the PPT criterion and the enhanced reduction criterion. That is, there are states that can be detected by one but not by the other. As so, given an unknown state, one may execute VED twice, one adopts the PPT criterion, and the other adopts the enhanced reduction criterion. 
The state is necessarily entangled if at least one of these two VEDs output "Entangled."
We also show by example how VED works in qutrit systems 
in Appendix~\ref{app:choi map}, utilizing the Choi map~\cite{Choi1975,choi1980some}.

\subsection{PPT criterion}\label{subsec:PPT}

A necessary condition for entanglement detection
is the positive partial transpose (PPT) criterion~\cite{peres1996separability},
which we briefly review as follows. Let $\rho_{AB}$ be a bipartite quantum state.
We can express it as
\begin{align}
    \rho_{AB} = \sum_{ijkl}\alpha_{ijkl}\ketbra{i}{j}_A\ox\ketbra{k}{l}_B,
\end{align}
where $\{\ket{i}\}_i$ and $\{\ket{k}\}_k$ are the computational bases of $A$ and $B$, respectively.
Its partial transpose with respect to system $B$ is defined as
\begin{align}
    \rho^{T_B}_{AB}
:=&\; (\id_A \ox T_B)(\rho_{AB}) \\
 =&\; \sum_{ijkl}\alpha_{ijkl}\ketbra{i}{j}\ox(\ketbra{k}{l})^T \\
 =&\; \sum_{ijkl}\alpha_{ijkl}\ketbra{i}{j}\ox\ketbra{l}{k},\label{eq:PPT}
\end{align}
where $T_B$ denotes the transpose map on system $B$.
The PPT criterion says that if $\rho_{AB}$ is separable,
then $\rho^{T_B}_{AB} \geq 0$.
Conversely, the negative spectrum witnesses entanglement of $\rho_{AB}$.
What's more, the PPT criterion is not only necessary but also sufficient
for separability of the $2\otimes 2$
and $2\otimes 3$ cases~\cite{stormer1963positive,woronowicz1976positive,horodecki1996Separability}.

We begin with the two-qubit bipartite quantum state case.
Notice that the qubit transpose map admits the following decomposition w.r.t. $\Delta$
specialized in~\eqref{eq:Delta-Pauli}:
\begin{align}\label{eq:T-decompose}
    T(\rho) = \frac{\rho + X \rho X - Y \rho Y + Z\rho Z}{2},
\end{align}
where $X, Y, Z$ are the Pauli matrices defined in~\eqref{eq:Pauli}.
The validity of this decomposition can be checked by direct calculation.
Substituting~\eqref{eq:T-decompose} into~\eqref{eq:PPT}, we obtain
\begin{align}
\rho^{T_B}_{AB}
:=&\; (\id_A\ox T_B)(\rho_{AB}) \\
=&\; \frac12(\rho + X_B\rho X_B - Y_B\rho Y_B + Z_B\rho Z_B),\label{eq:T-decompose-2}
\end{align}
where the quantum operation $X_B\rho_{AB} X_B$ should be understood as $(I_A\ox X_B)\rho_{AB} (I_A\ox X_B)$,
and similarly for $Y\rho Y$ and $Z\rho Z$.
Adapting the decomposition~\eqref{eq:T-decompose-2} into Algorithm~\ref{alg:VED},
we successfully apply the proposed VED to accomplish the PPT criterion in the qubit case.

Now we show the above detection method can 
be generalized to the multi-qubit bipartite quantum state case.
Let $\bm{B}\equiv B_1B_2\cdots B_n$ be a composite system with $n$
qubits, i.e., $B_i$ represents the $i$-th qubit system.
A key observation is that the transpose operation satisfies the tensor product 
property: transposing the composite system $\bm{B}$ is equivalent to 
transposing the local qubit systems $B_i$ individually. More precisely,
\begin{align}\label{eq:T-compose}
    T_{\bm{B}} = \bigotimes_{i=1}^n T_{B_i},
\end{align}
where $T_{B_i}$ is the transpose operation on the $i$-th qubit.
Eqs.~\eqref{eq:T-compose} and~\eqref{eq:T-decompose} together give $T_{\bm{B}}$ 
a linear combination into Pauli channels of $4^n$ terms in total.
Using this decomposition, we may apply VED (Algorithm~\ref{alg:VED} or 
Algorithm~\ref{alg:VED-via-sampling}) 
to accomplish the multi-qubit PPT criterion deterministically or probabilistically.

\subsection{Reduction criterion}\label{subsec:reduction}

In this section, we first review the reduction criterion~\cite{Cerf1999}
and then propose a variational algorithm implementing this criterion
within the VED framework described in~\ref{sec:VED}.
\begin{align}\label{eq:reduction map}
     \cR_{B\to B}(X_B) := \tr[X_B]I_B - X_B,
\end{align}
which is known as the \emph{reduction map}. The reduction criterion says that if a bipartite quantum state $\rho_{AB}$ is separable, then it must hold that
\begin{align}
\sigma_{AB} := \left(\id_A\ox\cR_{B\to B}\right)(\rho_{AB}) \geq 0.
\end{align}
Equivalently, if $\sigma_{AB}$ has negative eigenvalues, then $\rho_{AB}$
is entangled. It is based on this observation that our variational algorithm works. 

To apply the framework in~\ref{sec:VED}, we have first to decompose 
$\cR_{B\to B}$ into a linear combination of Pauli channels. Indeed, we can do so
since 
\begin{align}
\cR_{B\to B}(\rho_B) 
:=&\; \tr[\rho_B]I_B - \rho_B \\
 =&\; \frac{1}{2^n}\sum_{\bm{q}\in\{0,1,2,3\}^{\ox n}} 
        \cP_{\bm{q}}(\rho_B)I_B - \rho_B \\ 
 =&\; \frac{1-2^n}{2^n}\rho_B 
    + \frac{1}{2^n}\sum_{\bm{q}\neq\bm{0}} \cP_{\bm{q}}(\rho_B),\label{eq:reduction3}
\end{align}
where $\bm{0} \equiv (0,\cdots,0)$ of size $n$, and the second equality 
follows from the twirling property of Pauli channels~\cite[Exercise 4.7.3]{wilde2016quantum}.
Using this decomposition, we can call Algorithm~\ref{alg:VED} or Algorithm~\ref{alg:VED-via-sampling} 
to accomplish the reduction criterion.

Specially, in the qubit case where $n=1$, the reduction map is of the form
\begin{align}\label{eq:reduction4}
\cR_{B\to B}(\rho) = \frac{- \rho + X\rho X + Y\rho Y + Z\rho Z}{2}.
\end{align}

As one might see, deterministic VED using the reduction criterion
is not efficient in the multi-qubit case since it has to compute 
exponentially many numbers of overlaps: for a $2n$-qubit bipartite quantum state, one has to compute $4^n$ overlaps.
In Section~\ref{appx:improved-reduction}, we consider another version with better efficiency by exploring the simple structure of the reduction map~\eqref{eq:reduction map}.

On the other hand, probabilistic VED using the reduction criterion is also not efficient due to the sampling cost $\gamma\approx2^n$ for $\cR$. This observation leads to a simple way to improve the efficiency of the probabilistic VED.
For example, we introduce the trace-preserving reduction map, defined via
\begin{align}
    \widehat{\cR}_{B\to B}(\rho) := \frac{1}{2^n-1}\cR_{B\to B}(\rho_B).
\end{align}
One can check that as a positive map, $\widehat{\cR}$ has the same entanglement detection range 
as the original reduction map $\cR$. 
On the contrary, the sampling cost of $\widehat{\cR}$ is $\gamma=1+1/2^n$,
which is exponentially smaller than that of $\cR$.
This implies that probabilistic VED using the trace-preserving reduction 
map $\widehat{\cR}$ is very efficient in terms of the number of samples consumed.
However, the efficiency is achieved at the cost of the high precision required 
in estimating the minimal eigenvalue of $\widehat{\cR}(\rho_{AB})$,
which decreases exponentially in $n$ in general. We point out this is a trade off between efficiency and precision.
Similar arguments can be applied to the transpose map and the enhanced reduction map discussed in the next section.

\subsection{Enhanced reduction criterion}\label{subsec:enhanced-reduction}

In this section, we consider an enhanced version of the reduction
map~\cite{breuer2006optimal,hall2006new} for bipartite quantum states.
This enhanced criterion is based on an elementary positive map 
which operates on state spaces with \emph{even} dimension. 
It is known that the enhanced reduction criterion detects many bound entangled states (states that satisfy the PPT criterion). As before, we first review this enhanced reduction criterion and show how to combine it with the VED framework proposed in Sec.~\ref{sec:VED} to detect entanglement.

Define the following anti-symmetric unitary in a $n$-qubit Hilbert space:
\begin{align}
    U_a = \operatorname{antidiag}(1,-1,1,-1,\cdots,1,-1),
\end{align}
where $\operatorname{antidiag}$ means anti-diagonal. For example, 
when $n=2$, the corresponding anti-symmetric unitary has the form    
\begin{align}
    U_a = \begin{bmatrix}
            0 & 0 & 0 & 1 \\
            0 & 0 & -1 & 0 \\
            0 & 1 & 0 & 0 \\
            -1 & 0 & 0 & 0
        \end{bmatrix} = X \ox iY.
\end{align}
Indeed, one can check that the $n$-qubit $U_a$ 
can be decomposed w.r.t. the Pauli set as $U_a = X\ox\cdots\ox X\ox iY$,
where there are $n-1$ $X$ operators in the tensor product.
Based on $U_a$, we define the following map~\cite{breuer2006optimal}
\begin{align}\label{eq:enhanced-map}
\cK_{B\to B}(\rho_B) 
:= \cR_{B\to B}(\rho_B) - U_a T_B(\rho) U_a^\dagger,
\end{align}
where $\cR_{B\to B}$ is the reduction map defined in~\eqref{eq:reduction map}
and $T_B$ is the transpose map defined in~\eqref{eq:PPT}.
This map has been shown to be positive but not completely positive~\cite{breuer2006optimal}.
What's more, this map improves the reduction criterion 
and can detect bound entangled states that cannot be detected by the PPT criterion. 
Substituting the Pauli decomposition~\eqref{eq:reduction3} of $\cR$
and the Pauli decomposition~\eqref{eq:T-compose} of $T_B$ into~\eqref{eq:enhanced-map}
and regrouping the Pauli terms,
we obtain a Pauli decomposition of $\cK$, where there is a total number of $4^n$ Pauli terms.
Using on this decomposition, we can use VED (Algorithm~\ref{alg:VED} or Algorithm~\ref{alg:VED-via-sampling}) to accomplish the enhanced reduction criterion.

\subsection{Discussion on the measurement cost}
Note that another way to detect and quantify entanglement of a state $\rho_{AB}$ is to obtain its density matrix via quantum state tomography~\cite{nielsen2011quantum}. 
Full density matrix reconstruction of an unknown $(n_A+n_B)$-qubit state in the worst-case costs exponential copies of the state~\cite{ODonnell2016,Haah2017b}, e.g., $\widetilde \Omega(4^{n_A+n_B})$ measurement results are necessary to reconstruct a matrix close to $\rho$ in the sense of trace distance~\cite{Haah2017b}. Using the learned density matrix, we can either numerically apply a positive map on it or compute the fidelity between $\rho_{AB}$ and 
any entangled target states. However, such methods are resource-demanding compared to the VED framework. 

To use the decomposed maps for entanglement detection on the state $\rho_{AB}$, we can apply them either to subsystem $A$ or to subsystem $B$, which requires $poly(D)4^{\min\{n_A,n_B\}}$ measurement results with circuit depth $D$ in gradient-based optimization loops.
Besides the measurement cost, methods based on state tomography need vast memory to store and process the density matrix on a classical computer, which are resource-demanding as well. The VED framework, on the other hand, doesn't require such classical memory and post-processing.

\subsection{VED based on reduction criterion without decomposition}\label{appx:improved-reduction}

In Sec.~\ref{subsec:reduction} we have shown 
how VED uses the reduction criterion to detect entanglement; it works 
by decomposing the reduction map $\cR$ into a linear combination of Pauli channels and then variationally estimate the minimal eigenvalue of the averaged output state.

Here we propose another variational entanglement detection algorithm
for the reduction criterion, motivated by the simple structure of the reduction map.
The intuition behind this protocol is as follows. We know that $\rho_{AB}$ is entangled 
if $\cR_{B\to B}(\rho_{AB})$ is not semidefinite positive.
Using~\eqref{eq:min-eigenvalue}, this means that
\begin{align}
&\; \min_{\ket{\psi_{AB}}} \bra{\psi} \cR_{B\to B}(\rho_{AB}) \ket{\psi} \\
=&\; \min_{\ket{\psi_{AB}}} \bra\psi(I_A\otimes\rho_{B} -\rho_{AB})\ket\psi \\
=&\; \min_{\ket{\psi_{AB}}} \left\{\tr[\psi_B\rho_B] - \tr[\psi_{AB}\rho_{AB}]\right\}
< 0,\label{eq:appx:reduction3}
\end{align}
where the minimization ranges over all pure bipartite quantum states $\ket{\psi_{AB}}$
in system $AB$, $\psi_{AB}\equiv\proj{\psi}_{AB}$ and $\psi_B :=\tr_A\psi_{AB}$.
From Eq.~\eqref{eq:appx:reduction3}, one can see that it suffices
to compute the difference of two overlaps and then variationally estimate
the minimal eigenvalue. The crucial point is that the number of overlaps 
is independent on the dimension of the $n$-qubit system $B$.
This new detection method could save a large amount of computing resources 
when $n$ becomes large.
The improved VED based on the reduction criterion is summarized in
Algorithm~\ref{alg:reduction-alt}.

\begin{algorithm}[H] 
\caption{Improved VED based on reduction criterion}
\begin{algorithmic}[1] \label{alg:reduction-alt}

\STATE Input: $2n$-qubit quantum state $\rho_{AB}$, 
        parameterized quantum circuits $U(\bm\a)$ with initial parameters $\bm\a$, and
        tolerance $\delta$;
    
\STATE  Apply $U(\bm\a)$ to $\ket{00}_{AB}$ on system $AB$ and 
        obtain the test state $\ket {\psi}_{AB} = U(\bm\a)\ket{00}_{AB}$; 
 
\STATE  Compute the overlap between state $\psi_B$ and $\rho_B$ 
        on subsystem $B$ using the Swap Test and obtain $c_1 = \tr[\psi_B\rho_B]$;

\STATE  Apply $U(\bm\a)$ to $\ket{00}_{AB}$ on system $AB$ and 
        obtain the test state $\ket{\psi}_{AB} = U(\bm\a)\ket{00}_{AB}$; 

\STATE  Compute the overlap between state $\psi_{AB}$ and $\rho_{AB}$ 
        using the Swap Test and obtain $c_2 = \tr[\psi_{AB}\rho_{AB}]$;

\STATE Compute the loss function $L(\bm\a) =  c_1 - c_2$;

\STATE Perform optimization methods to minimize $L(\bm\a)$; 
        terminate the optimization if  
        the error tolerance is satisfied: $L(\bm\a) < -\delta$.

\STATE Output "Entangled" if the optimized $L(\bm\a) < -\delta$.
\end{algorithmic}
\end{algorithm}
We remark that this idea can also be adopted to 
improve the efficiency of VED using the enhanced reduction criterion.

\section{Quantum entanglement quantification}\label{sec:VEQ}

One of the most well-known entanglement measure is the logarithmic negativity~\cite{Vidal2002,Plenio2005b}, 
which has various applications 
in quantum information theory. 
For a bipartite state $\rho_{AB}$, 
its logarithmic negativity is defined as
\begin{align}
    E_N(\rho_{AB}):=\log \|\rho_{AB}^{T_B}\|_1.
\end{align}
Based on the recently developed near-term quantum algorithm for trace distance estimation~\cite{Chen2020a}
and the fact that $E_N$ is defined via the transpose map $T_B$,
we introduce a variational quantum algorithm to estimate $E_N$
using an ancillary qubit system $R$. According to~\cite[Corollary 3]{Chen2020a}, it holds that
\begin{align}
    \|\rho_{AB}^{T_B}\|_1 &= 2\max_U\tr\proj{0}_R Q_R - \tr \rho_{AB}^{T_B}\\
    &= 2\max_U\tr\proj{0}_R Q_R - 1,\label{eq:VEQ2}
\end{align}
where $Q_R = \tr_{AB} Q_{ABR}$, $Q_{ABR} = U(\rho_{AB}^{T_B}\otimes\proj{0}_R)U^\dagger$, 
and the maximization ranges over all unitaries on the composite system $ABR$. 
Note that the second equality follows from the fact that $T_B$ is trace-preserving.
Following the idea of VED, we may decompose
the transpose map $T_B$ appeared in the operator $Q_{ABR}$ (correspondingly, $Q_R$)
into a linear combination of Pauli terms via~\eqref{eq:T-decompose-2} and~\eqref{eq:T-compose},
compute the overlaps in~\eqref{eq:VEQ2} one by one, and then variationally estimate the maximal value. 
For illustrative purposes, we give Algorithm~\ref{alg:vtde}, 
the \emph{Variational Logarithmic Negativity Estimation (VLNE)}, 
as an example of estimating the logarithmic negativity of a two-qubit quantum state $\rho_{AB}$. 
However, we emphasize that method outlined in Algorithm~\ref{alg:vtde} can be easily 
generalized to quantify multi-qubit bipartite entanglement, 
as the transpose operation satisfies the 
preferable tensor product property~\eqref{eq:T-compose}.
What's more, Algorithm~\ref{alg:vtde} can be modified to
use the sampling technique to estimate the average state,
following the idea illustrated in
Algorithm~\ref{alg:VED-via-sampling}.

\vspace{1cm}
\begin{algorithm}[H]
\caption{Variational Logarithmic Negativity Estimation}
\begin{algorithmic}[1] \label{alg:vtde}
\STATE Input: a $2$-qubit quantum state $\rho_{AB}$ and parameterized circuits $U_{ABR}(\bm\a)$ with initial parameters $\bm\a$;

\STATE Apply $U_{ABR}(\bm\a)$ respectively to 
\begin{align}
&\rho_{AB}\otimes\proj{0}_R,\\
&(I_A\otimes X_B)\rho_{AB}(I_A\otimes X_B)\otimes\proj{0}_R,\\
&(I_A\otimes Y_B)\rho_{AB}(I_A\otimes Y_B)\otimes\proj{0}_R,\\
&(I_A\otimes Z_B)\rho_{AB}(I_A\otimes Z_B)\otimes\proj{0}_R, 
\end{align}
and obtain the states $\sigma^{(0)}$, $\sigma^{(1)}$, $\sigma^{(2)}$, $\sigma^{(3)}$, respectively.
 
\STATE Obtain $o_j=\tr[\sigma^{(j)}_R\proj{0}_R]$ for $j=0,1,2,3$ by measurements on system $R$. 

\STATE Compute the loss function $\cL_1:=-(o_0+o_1-o_2+o_3)/2$.
 
\STATE Perform optimization methods to minimize $\cL_1(\bm\a)$;

\STATE Compute $\b = 2|\cL_1|-1$ as the estimated trace norm of $\rho_{AB}^{T_B}$;

\STATE Output $\log\b$ as the estimated logarithmic negativity.
\end{algorithmic}
\end{algorithm}

One may also evaluate the entanglement 
measures~\cite{chen2014comparison,zhu2017coherence,wang2019quantifying}
based on the sandwiched R\'{e}nyi relative entropy~\cite{mueller-lennert2013quantum,wilde2014strong}
of order $1/2$, making use of the recently proposed variational 
quantum algorithm estimating the fidelity between two quantum states~\cite{Chen2020a}.

\section{Experiments in IBMQ}
\label{sec:ibmq-santiago}

\begin{figure}[t]
\[\Qcircuit @C=0.8em @R=0.5em{
\lstick{\ket{0}} &\gate{R_y(\a_1)} &\gate{R_z(\a_3)} &\qw &\ctrl{1}  &\qw
&&& \raisebox{-9ex}{  $\Bigg \} \,  \ket{\psi} $  } \\
\lstick{\ket{0}} &\gate{R_y(\a_2)} &\qw              &\qw &\targ  &\qw
}\]
\caption{Parameterized two-qubit quantum circuit $U(\bm\alpha)$ used for 
preparing the test state $\psi_{AB}(\bm \a)$  on the ibmq-santiago hardware. 
The parameters $\bm \a$ are randomly initialized as $(\a_1,\a_2,\a_3)= (3.2292,4.8579,5.4691)$.}
\label{fig:ibmq-ansatz}
\end{figure}
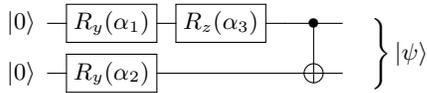

In this section, we discuss how to apply the VED framework to detect the two-qubit 
maximally entangled state $\ket{\Phi}:=(\ket{00}+\ket{11})/\sqrt{2}$ on IBM-Q superconducting 
quantum hardware accessible to the public. The specific quantum device used is ibmq-santiago (5 qubits) with a quantum volume of $32$.
The positive map adopted here for detection purpose is the qubit reduction map $\cR_{B\to B}$ defined in Eq.~\eqref{eq:reduction map}. 
After implementing the decomposed reduction map by $4$ Pauli terms as Eq.~\eqref{eq:reduction4}, 
we use a parametrized quantum circuit $U(\bm \a)$ to prepare $4$ identical test 
states $\psi_{AB}(\bm \a) = U(\bm \a)\proj{00}_{AB}U^\dagger(\bm\a)$ and 
compute the loss function defined in Eq.~\eqref{eq:loss}. 
The PQC used is depicted in Fig.~\ref{fig:ibmq-ansatz} 
with three randomly initialized parameters $\bm\a = (\a_1,\a_2,\a_3)$. 
During the optimization procedure, we apply the gradient descent algorithm~\cite{lemarechal2012cauchy} to guide the learning process where the analytical gradient is calculated via the following parameter-shift rule~\cite{mitarai2018quantum}:
\begin{align}
\frac{\partial \cL(\bm \a)}{\partial \a_j} 
:= \frac{1}{2} \left[ \cL\left(\a_j+\frac{\pi}{2}\right) - 
                        \cL\left(\a_j-\frac{\pi}{2}\right)\right].
\end{align}

\begin{figure}[h]
    \centering
    \includegraphics[width=0.5\textwidth]{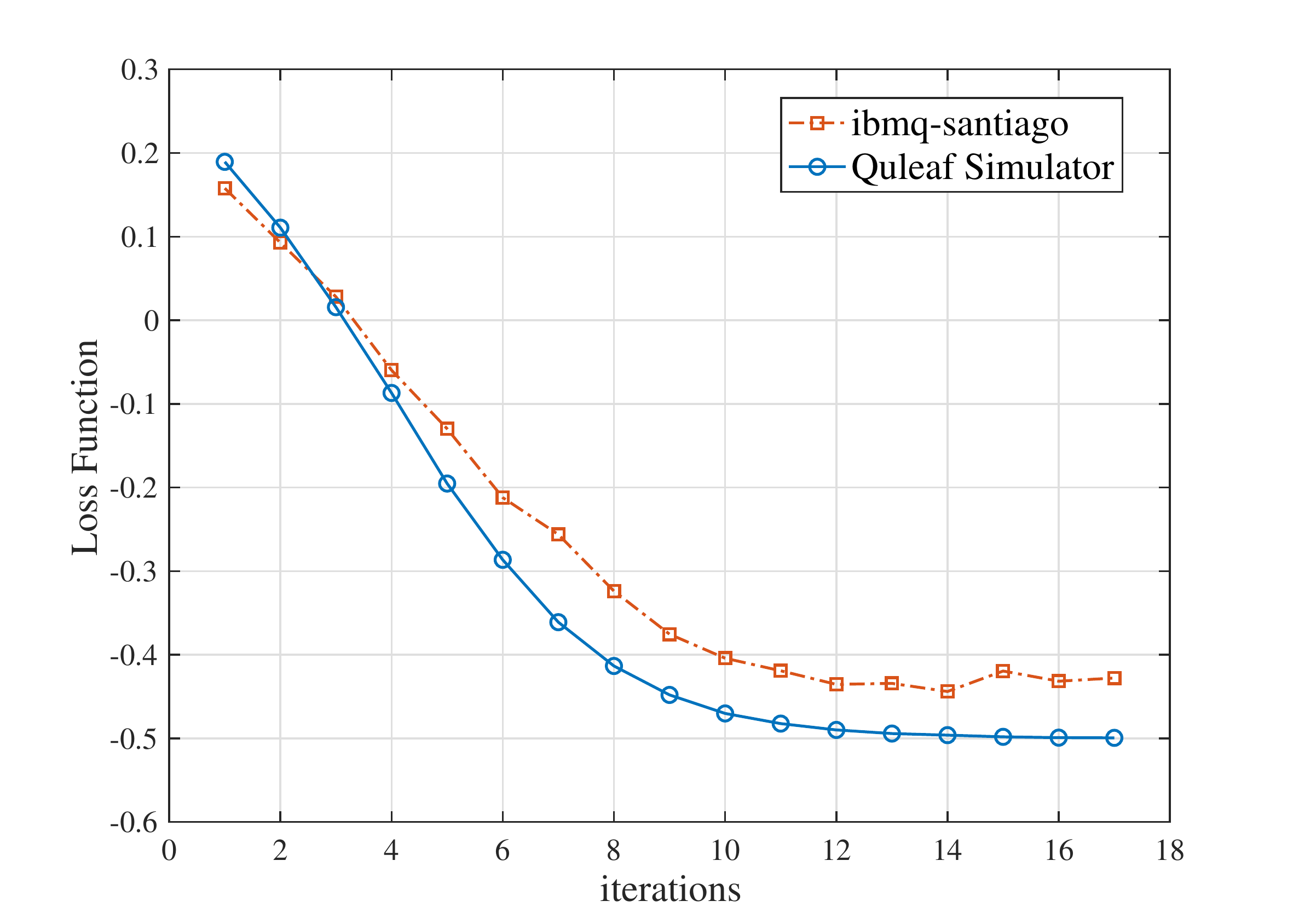}
    \caption{Estimated $\lambda_{\text{min}}$ by VED using the reduction criterion on the 
        Bell state $\ket{\Phi}$. 
        The red curve records the results from
        ibmq-santiago with shots = $8192$ for each circuit evaluation. 
        The blue curve records the simulation results on  \href{https://quantum-hub.baidu.com/}{Quantum Leaf} platform using the Baidu Quleaf simulator~\cite{Quleaf}.
        Learning rate in the gradient descent algorithm is set to be LR $=0.5$.}
    \label{fig:ibmq-res}
\end{figure}

Due to the finite sampling restriction for measurements, the optimization procedure essentially falls into the regime of Stochastic Gradient Descent (SGD)~\cite{kiefer1952stochastic}. The optimized loss values converges to $\cL_{\min} \approx -0.43$. The gap between the experiment data and simulation result $\lambda_{\text{min}} = -0.5$ is due to various hardware noises on the ibmq-santiago processor. One can further adopt error mitigation methods~\cite{Endo2020} to improve the result.
This result proves the validity of our VED framework. Note that if we adopt the termination setup in Algorithm~\ref{alg:VED}, it will require much fewer optimization iterations ($4$ to $5$ rounds are sufficient) to obtain the detection result. 
As mentioned in Ref.~\cite{coles2018quantum}, the communication bottleneck between the IBM-Q hardware and classical optimizer blocks us from efficiently conducting experiments without any specified reservation. This leads to a $9$-minutes waiting time on average for each circuit evaluation from the IBM-Q cloud service.
As a comparison, we conduct numerical simulations on 
\href{https://quantum-hub.baidu.com/}{Quantum Leaf} platform and receive a similar result by repeating the simulation on qiskit-Aer simulator. We summarize the experimental and numerical results in Fig.~\ref{fig:ibmq-res}. 

\section{Numerical simulations}
\label{sec:numercs}

In this section, we carry out numerical simulations that apply VED to detect a variety of bipartite quantum states of interest to investigate
the performance of VED and its motivated entanglement quantification algorithm. All simulations, including optimization loops, are conducted 
using the Paddle Quantum~\cite{Paddlequantum} toolkit on the PaddlePaddle Deep Learning
Platform~\cite{Paddle,Ma2019}.

\subsection{Entanglement detection}

\begin{figure}[!hbtp]
\[\Qcircuit @C=0.8em @R=0.5em{
&\gate{U_3(\t_{0,0},\t_{0,1},\t_{0,2})}&\ctrl{1} &\qw &\targ  &\qw  &\cdots &\quad &\gate{U_3(\t_{2,0},\t_{2,1},\t_{2,2})} &\qw\\ 
&\gate{U_3(\t_{0,3},\t_{0,4},\t_{0,5})}&\targ &\ctrl{1}  &\qw  &\qw  &\cdots &\quad &\gate{U_3(\t_{2,3},\t_{2,4},\t_{2,5})} &\qw\\
&\gate{U_3(\t_{0,6},\t_{0,7},\t_{0,8})}&\qw &\targ     &\ctrl{-2} &\qw^{\quad \quad \quad \,\times 2}  &\cdots &\quad &\gate{U_3(\t_{2,6},\t_{2,7},\t_{2,8})} &\qw
\gategroup{1}{2}{3}{5}{2.07em}{--}
}\]
\caption{Three qubit parameterized ansatz $U(\bm\alpha)$ used for VED. 
The quantum circuit within the dotted block is repeated twice.}
\label{fig:ansatz}
\end{figure}
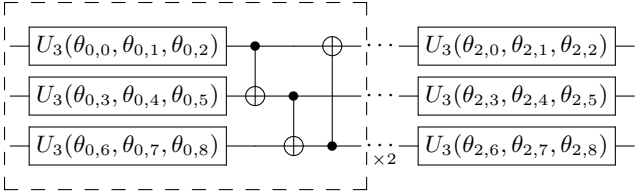

For the entanglement detection purpose, we adopt the circuit ansatz shown in Fig.~\ref{fig:ansatz}
to prepare the test state $\ket{\psi}_{AB}$. 
It consists of parameterized single-qubit gates $U_3(\t,\phi,\varphi) = R_z(\phi)R_y(\t)R_z(\varphi)$ 
and circular layers of CNOT gates. Note that this ansatz can be easily generalized to multi-qubit case.

\subsubsection{Four-qubit isotropic states}
The four-qubit isotropic state family is defined as~\cite[Eq. (32)]{horodecki1999reduction}
\begin{align}\label{eq:isotropic}
    \rho^{\text{iso}}_{\bm{A}\bm{B}}(p) := p\Phi_{\bm{AB}} + (1-p)\frac{I_{\bm{AB}}}{16},
\end{align}
where $p\in[0,1]$ is a parameter, $\Phi_{\bm{AB}}$ is the four-qubit maximally entangled state, and $I_{\bm{AB}}$ is the identity operator in $\bm{AB}$ in which $\bm{A}\equiv A_1A_2$ and $\bm{B}\equiv B_1B_2$.
The qubit systems $A_1$ and $A_2$ are at Alice's hand, while the qubit systems $B_1$ and $B_2$ are at Bob's hand.
Intuitively, the isotropic state is a convex combination of the maximally entangled state $\Phi_{\bm{AB}}$ and the maximally mixed state $I_{\bm{AB}}/16$.
It has been shown that $\rho^{\text{iso}}_{\bm{AB}}(p)$ is separable (w.r.t. the $\bm{A}{:}\bm{B}$ cut) if and only if $p\leq1/5$~\cite{horodecki1999reduction}.

\begin{figure}[!hbtp]
    \centering
    \includegraphics[width=0.5\textwidth]{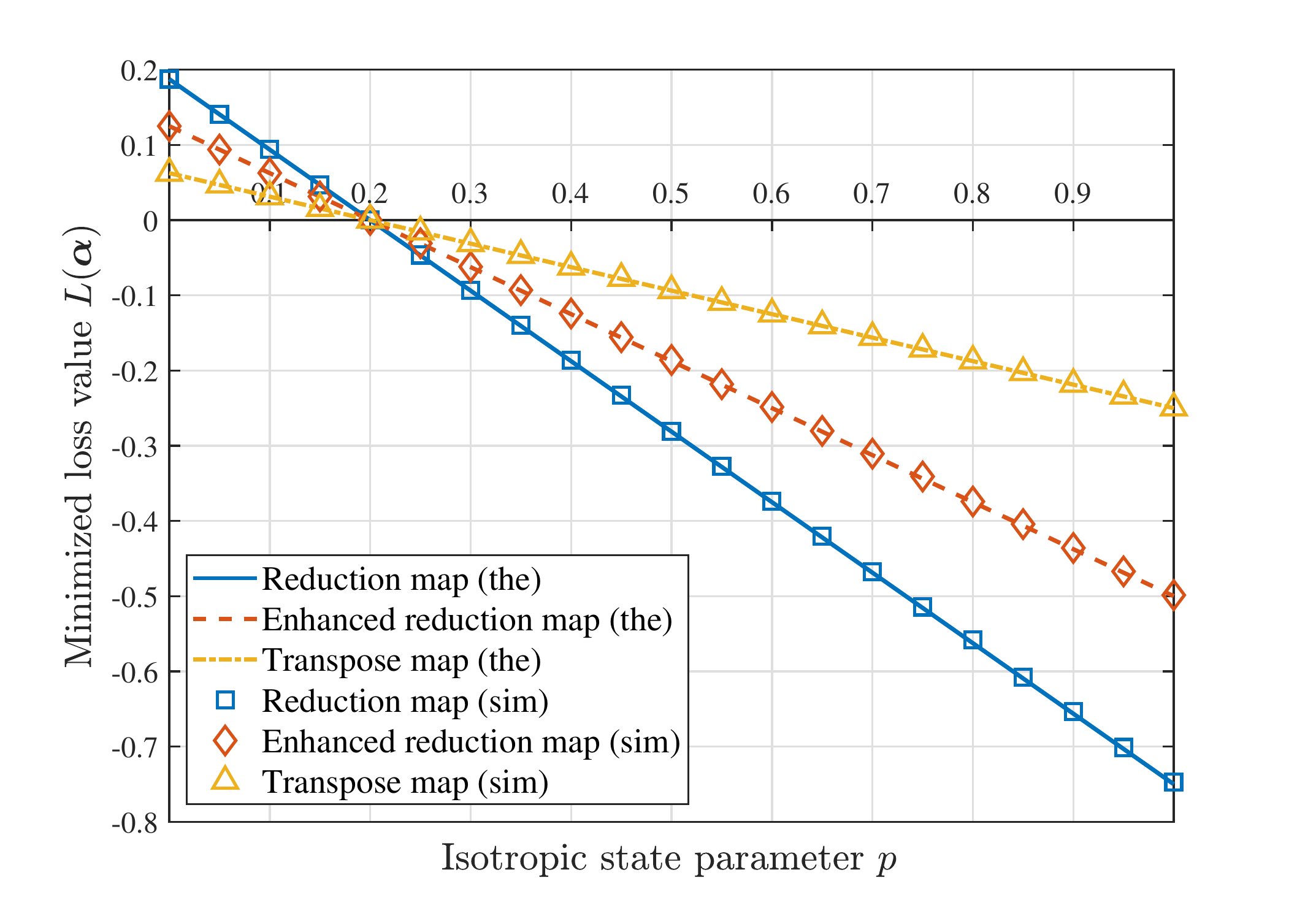}
    \caption{Numerical results on the four-qubit isotropic states defined in~\eqref{eq:isotropic}. Each line depicts the smallest eigenvalue of every isotropic state with parameter $p\in[0,1]$ under the corresponding map. This line of the smallest eigenvalues is a lower bound of the loss function $L(\bm\a)$. Each marker depicts the minimized loss value obtained by simulations (sim) of Algorithm~\ref{alg:VED} on a chosen isotropic state, aligning with the theoretical line.}
    \label{fig:isotropic4}
\end{figure}

We numerically carry out Algorithm~\ref{alg:VED} together with the three prominent positive maps---the PPT criterion, the reduction criterion, and the enhanced reduction criterion---introduced in Sec.~\ref{sec:pm}, using four-qubit isotropic states as inputs. The minimized loss values of these three maps obtained by our simulations on the isotropic states are represented by different markers in Fig.~\ref{fig:isotropic4}.
As can be seen from this figure, VED can successfully identify the range of $p$ for which the corresponding isotropic state can be detected by each positive map. The markers representing results from simulations fall on the lines that give the minimums of the loss function $L(\bm\a)$, verifying the validity and viability of our VED framework. Note that for detecting entanglement in four-qubit isotropic states, all three maps are both necessary and sufficient. However, this phenomenon is not universal for all four-qubit states, as we shall see in the experiment using Breuer states.

\subsubsection{Four-qubit Breuer states}

As we have mentioned in Sec.~\ref{subsec:enhanced-reduction}, there are states 
that can be detected by the enhanced reduction criterion yet cannot be 
detected by the PPT criterion. In this section, we use the proposed VED framework
to numerically consolidate this statement.
The four-qubit Breuer state family is
defined as~\cite[Eq. (7)]{breuer2006optimal}
\begin{align}\label{eq:Breuer}
\rho^{\text{Breuer}}_{\bm{A}\bm{B}}(\lambda) := 
\begin{pmatrix}
    \frac{1-\lambda}{3} & 0 & 0 & 0 \\
    0 & \frac{1+2\lambda}{6} & \frac{1-4\lambda}{6} & 0 \\
    0 & \frac{1-4\lambda}{6} & \frac{1+2\lambda}{6} & 0 \\
    0 & 0 & 0 & \frac{1-\lambda}{3} 
\end{pmatrix},
\end{align}
where $\lambda\in[0,1]$ is a parameter, $\bm{A}\equiv A_1A_2$, and $\bm{B}\equiv B_1B_2$. 
The qubit systems $A_1$ and $A_2$ are at Alice's hand while 
the qubit systems $B_1$ and $B_2$ are at Bob's hand.
It has been shown that $\rho^{\text{Breuer}}$
is separable (w.r.t. the $\bm{A}{:}\bm{B}$ cut) if and only if $\lambda=0$ 
and can be detected by the enhanced reduction criterion~\cite{breuer2006optimal}.
On the other hand, it has positive partial transpose 
if and only if $\lambda\leq1/6$~\cite{breuer2006optimal}, 
witnessing the power of the enhanced reduction criterion. 

\begin{figure}[!hbtp]
    \centering
    \includegraphics[width=0.5\textwidth]{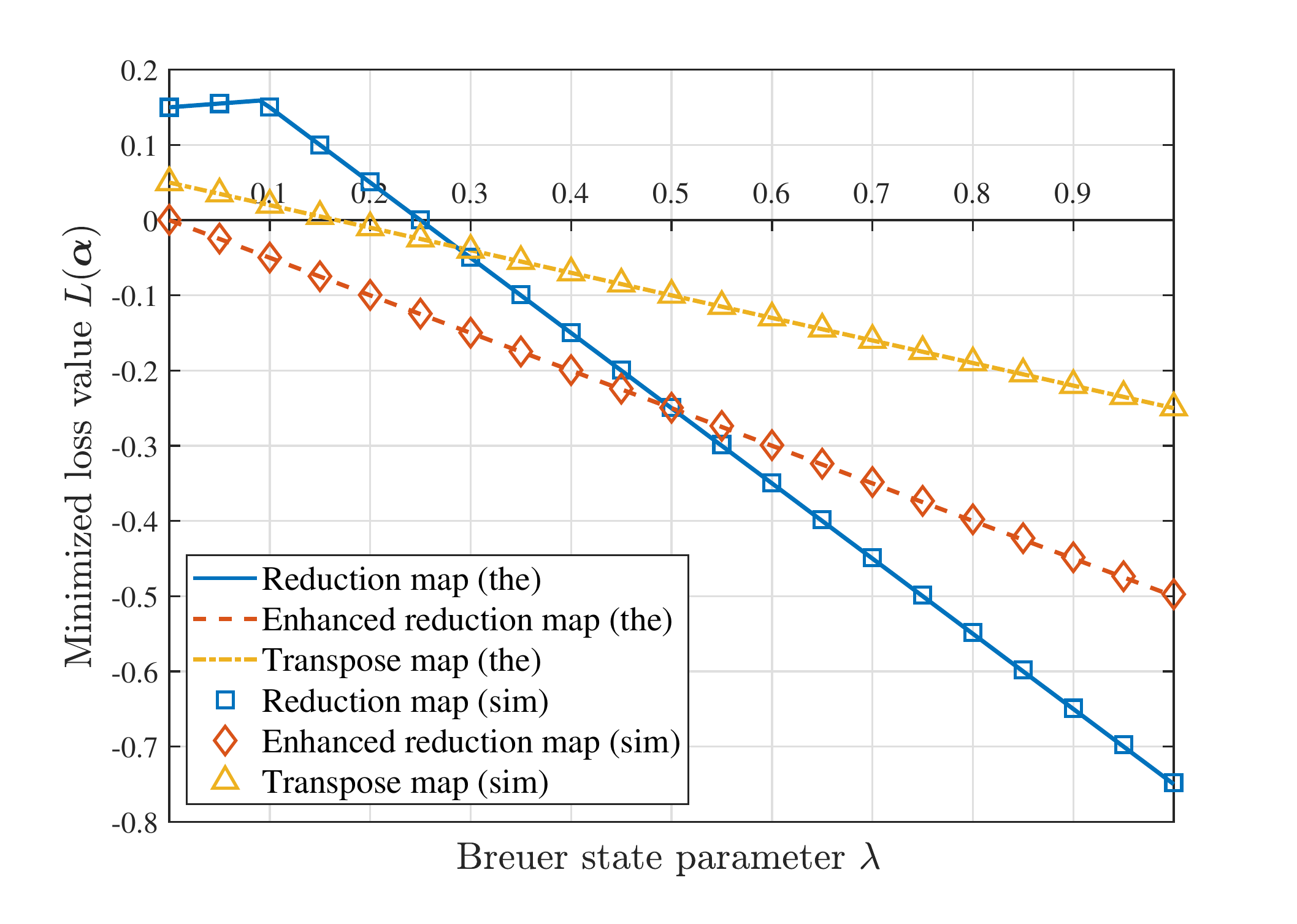}
    \caption{Numerical results on the four-qubit 
    Breuer states defined in~\eqref{eq:Breuer}. 
    Each line depicts the smallest eigenvalue of every
    Breuer state with parameter $p\in[0,1]$ under the corresponding map.
    This line of the smallest eigenvalues is a lower bound of the loss function $L(\bm\a)$.
    Each marker depicts the minimized loss value obtained by 
    simulations (sim) of Algorithm~\ref{alg:VED} on a chosen Breuer state,
    aligning with the theoretical line.}
    \label{fig:Breuer}
\end{figure}

Following the same line of the case of the isotropic state,
we carry out Algorithm~\ref{alg:VED} on the
three criteria using four-qubit Breuer states as inputs. The minimized loss values obtained by our simulations on selected Breuer states are represented in Fig.~\ref{fig:Breuer} by markers,  which again aligns with the theoretical lines. From the numeric results, we can see that while the enhanced reduction criterion is sill necessary and sufficient for entanglement detection in the four-qubit Breuer states, neither the reduction criterion nor the PPT criterion can detect all entangled states in the Breuer state family, attesting the advantage of the enhanced reduction criterion in this case.

\subsection{Logarithmic negativity estimation}

\begin{figure}[t]
    \centering
    \includegraphics[width=0.5\textwidth]{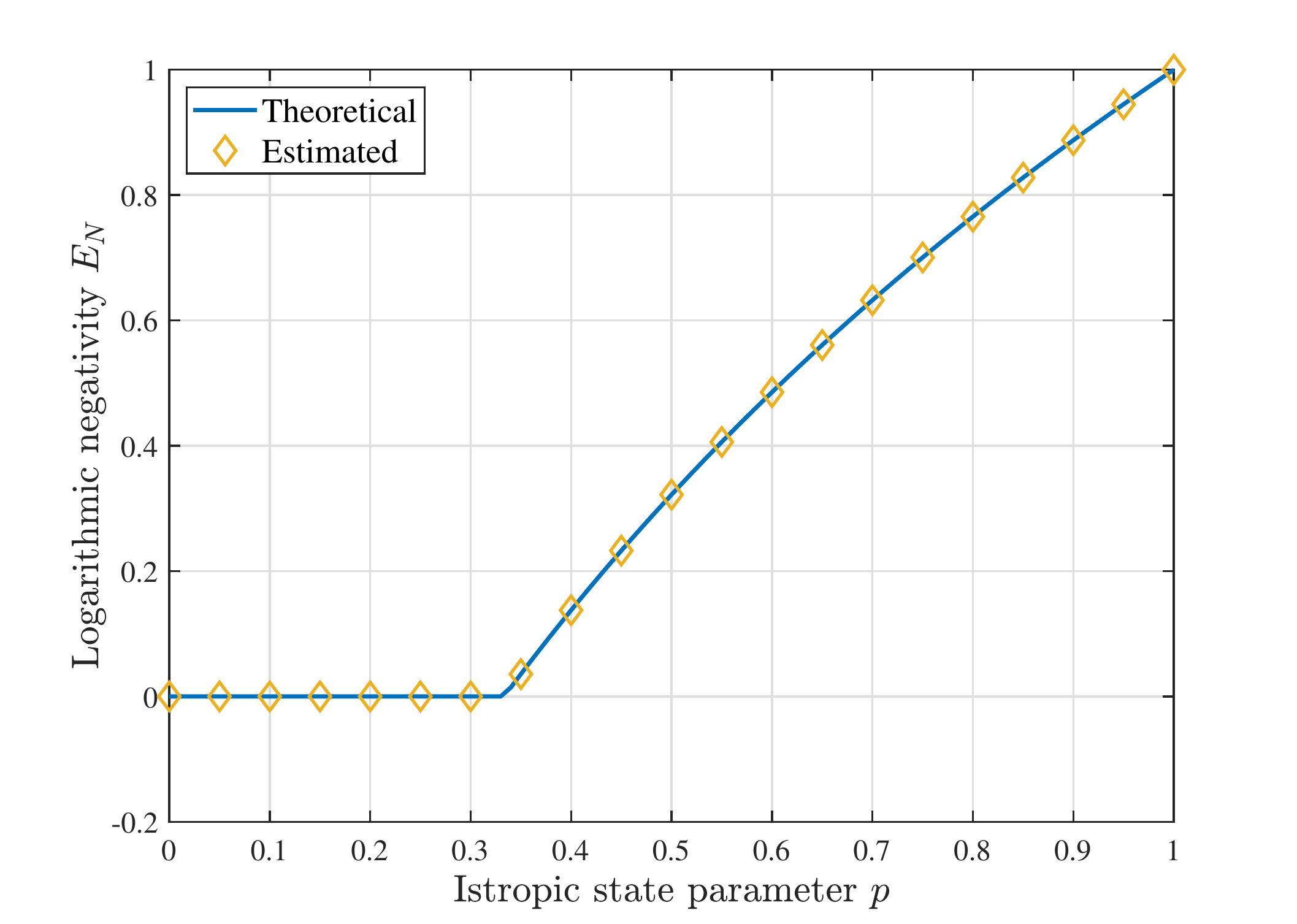}
    \caption{Numerical results on the two-qubit isotropic states. The blue line represents the precisely calculated logarithmic negativity of isotropic states with parameter $p\in[0,1]$. The yellow markers depict the estimated  logarithmic negativity by simulations of Algorithm~\ref{alg:vtde} on selected isotropic states.}
    \label{fig:iso_logneg}
\end{figure}

For simulations of variational entanglement quantification with logarithmic negativity, 
we adopt the hardware efficient ansatz used for trace distance estimation in~\cite{Chen2020a} where the circuit
depth is $4$. The simulations are carried out on two-qubit isotropic states, which is defined as
\begin{align}\label{eq:isotropic_2q}
    \rho^{\text{iso}}_{AB}(p) := p\Phi_{AB} + (1-p)\frac{I_{AB}}{4},
\end{align}
where $\Phi_{AB}$ is the two-qubit maximally entangled state.
As shown in Fig.~\ref{fig:iso_logneg}, the logarithmic negativity of a two-qubit isotropic state is positive if and only if its parameter $p > 1/3$, which matches the range of $p$ where the corresponding isotropic states are entangled. The estimated logarithmic negativities by our method, which are represented by markers in Fig.~\ref{fig:iso_logneg}, agree with the precisely calculated values given by the blue line. 

\section{Conclusion and outlook}\label{sec:conclusion}

In this work, we combined two novel techniques that find crucial applications in the NISQ quantum devices, the variational quantum algorithms and the quasi-probability decomposition method, to propose the Variational Entanglement Detection (VED) and Variational Logarithmic Negativity Estimation (VLNE) frameworks, contributing feasible solutions to detect and quantify entanglement on near-term devices. VED is built upon the positive map criterion and works as follows. Firstly, it decomposes a chosen positive map into a linear combination of NISQ implementable quantum operations. Then, it variationally estimates the minimal eigenvalue of the output state of some positive map acting on the target bipartite state. Two methods are proposed to generate the output state:
the first one averaged the output states according to the quasi-probability distribution; 
the second one estimated the average via the sampling technique.
At last, it asserts that the target state is entangled if the optimized minimal eigenvalue is negative, guaranteed by the positive map criterion. We elaborated three well-known positive maps to illustrate how the VED framework is applied. Following the idea of VED, VLNE variationally computes the log-negativity entanglement measure,  relying on a linear decomposition of the transpose map into Pauli terms and the recently proposed trace distance estimation algorithm. Experimental and numerical results on various bipartite states of interest have validated the proposed entanglement detection and quantification methods.

We expect that the VED framework can be upgraded to detect more entangled states. A crucial step towards this aim is to explore what kind of positive maps can be decomposed into a linear combination of Pauli channels. In Sec.~\ref{sec:VEQ} we showed by case how to variationally compute the log-negativity entanglement measure. It would be meaningful to design novel quantum algorithms to estimate other distance-based entanglement measures~\cite{vedral1997quantifying,vedral1998entanglement,rains2001semidefinite,datta2009min,zhu2017coherence}.

\section*{Acknowledgements}
We thank Runyao Duan for helpful suggestions.  XW would like to thank Youle Wang for useful discussions.

\bibliographystyle{apsrev4-1}
\bibliography{smallbib}

\appendix

\section{Qutrit entanglement detection via the Choi map}\label{app:choi map}

In this Appendix, we show by example how VED
can be adapted to detect the entanglement of an unknown two-qutrit state
via the celebrated Choi map~\cite{Choi1975,choi1980some}. 
To be specific, the Choi map is defined as follows
\begin{equation}
\Phi_{C}(\rho) :=
\begin{bmatrix}
    \rho_{1,1}+\rho_{2,2} & - \rho_{1,2} & -\rho_{1,3} \\
    -\rho_{2,1} & \rho_{2,2}+\rho_{3,3} & -\rho_{2,3} \\
    -\rho_{3,1} & -\rho_{3,2} & \rho_{3,3}+\rho_{1,1}
\end{bmatrix},
\end{equation}
where $\rho_{i,j}$ is the element of $\rho$ in the $i$-th row and $j$-th column.
This map is the first known  example of a positive map that is indecomposable (i.e., it cannot be decomposed into a sum of a completely positive map and a completely copositive map).
The Choi map in particular can be used to detect entanglement 
of some PPT states (states that can not be detected by the transpose map).

For the qutrit system, the generalized Pauli matrices are defined through 
the unitary boost and shift operators 
\begin{align}
X=
\begin{bmatrix}
0 & 0 & 1  \\
1 & 0 & 0  \\
0 & 1 & 0
\end{bmatrix}
\quad \text{and}\quad 
Z=
\begin{bmatrix}
1 & 0 & 0  \\
0 & \omega & 0  \\
0 & 0 & \omega^{2}
\end{bmatrix},
\end{align}
where $\omega:=e^{2\pi i/3}$ is the $3$-th root of unity.
The Choi map can be decomposed w.r.t. $X$ and $Z$ as follows:
\begin{align} 
\Phi_{C}(\rho) 
&= \frac{1}{3}[X^2Z\rho (X^2Z)^\dagger + X^2Z^2\rho (X^2Z^2)^\dagger + X^2\rho (X^2)^\dagger] \nonumber \\
&\qquad + \frac{2}3[Z\rho Z^\dagger +Z^2\rho (Z^2)^\dagger]-\frac{1}{3}\rho.
    \label{eq:choi decom}
\end{align}

Based on~\eqref{eq:choi decom}, one could adopt 
the deterministic VED in Algorithm \ref{alg:VED} 
or the probabilistic VED in Algorithm \ref{alg:VED-via-sampling} 
to detect entanglement on qutrit systems by considering
\begin{align}
    \min_{\psi(\bm\a)}\bra{\psi(\bm\a)} \Phi_{C}(\rho_{AB})\ket{\psi(\bm\a)},
\end{align}
where the test state $\psi(\bm\a)$ can be generated by parameterized quantum circuits.

\end{document}